\documentclass[twocolumn, showpacs,preprintnumbers,amsmath,amssymb]{revtex4}
\usepackage{enumitem}
\usepackage{amssymb}
\usepackage{xcolor}
\usepackage{graphicx}
\usepackage{dcolumn}
\usepackage{bm}
\usepackage{multirow}
\usepackage{booktabs}
\usepackage{longtable}
\usepackage{natbib}
\usepackage{lscape}
\usepackage{soul}
\usepackage{amssymb}
\usepackage{amsmath}	
\usepackage{color}
\usepackage{cases} 

\usepackage{array}
\newcommand{\Kmat}{\mbox{\boldmath $\mathcal{K}$}}

\newcommand{\Vmat}{\mbox{\boldmath $\mathcal{V}$}}

\newcommand{\Umat}{\mbox{\boldmath $\mathcal{U}$}}
\newcommand{\etamat}{\mbox{\boldmath $\eta$}}
\newcommand{\numat}{\mbox{\boldmath $\nu$}}
\newcommand{\Cmat}{\mbox{\boldmath $\cal C$}}
\newcommand{\Smat}{\mbox{\boldmath $\cal S$}}
\newcommand{\SSmat}{\mbox{ $S$}}
\newcommand{\Xmat}{\mbox{\boldmath $X$}}

\begin{document}
\preprint{APS/123-QED}

\title{Rotational transitions induced by collisions of HD$^{+}$ ions with low energy electrons}
\author{O. Motapon$^{1,2}$}
\author{N. Pop$^{3}$}
\author{F. Argoubi$^{4}$}
\author{J. Zs Mezei$^{2,5,6}$}
\author{M. D. Ep\'ee Ep\'ee$^{1}$}
\author{A. Faure$^{7}$}
\author{M. Telmini$^{4}$}
\author{J. Tennyson$^{8}$}
\author{I. F. Schneider$^{2,5}$}\email[]{ioan.schneider@univ-lehavre.fr}
\affiliation{$^{1}$LPF, UFD Math., Info. Appliq. Phys. Fondamentale, University of Douala, P. O. Box 24157, Douala, Cameroon}%
\affiliation{$^{2}$LOMC CNRS$-$Universit{\'{e}} du Havre$-$Normandie Universit{\'{e}}, 76058 Le Havre, France}
\affiliation{$^{3}$Dept. of Physical Foundation of Engineering, University Politechnica of Timisoara, 300223, Timisoara, Romania}%
\affiliation{$^{4}$LSAMA, University of Tunis El Manar, 2092 Tunis, Tunisia}%
\affiliation{$^{5}$LAC, CNRS$-$Universit\'e Paris-Sud$-$ENS Cachan$-$Universit\'e Paris-Saclay, 91405 Orsay, France}%
\affiliation{$^{6}$HUN-REN Institute for Nuclear Research (ATOMKI), H-4001 Debrecen, Hungary}%
\affiliation{$^{7}$IPAG, CNRS-INSU-Universit\'e UJF-Grenoble 1, 38000 Grenoble, France}
\affiliation{$^{8}$Dept. of Physics and Astronomy, University College London, WC1E 6BT London, UK}%
\date{\today}

\begin{abstract}
A series of Multichannel Quantum
Defect Theory-based computations have
been performed, in order to produce
 the cross sections of rotational transitions
(excitations $N_{i}^{+}-2 \rightarrow$  $N_{i}^{+}$, de-excitations
$N_{i}^{+}$ $\rightarrow$ $N_{i}^{+}-2$, with $N_{i}^{+}=2$ to $10$) and
of their competitive process, the dissociative recombination, induced by
collisions of HD$^+$ ions with electrons in the energy
range $10^{-5}$ to 0.3 eV. Maxwell anisotropic rate coefficients, obtained from
these cross sections in
the conditions of the Heidelberg Test Storage Ring (TSR) experiments
($k_{B}T_{t}=2.8$ meV and $k_{B}T_{l}=45$ $\mu$eV), have been reported for those processes in
the same electronic energy range. Maxwell isotropic rate coefficients have
been as well presented for electronic temperatures up to a few hundreds of Kelvins.
Very good overall agreement is found between our results for rotational transitions and the former
theoretical computations as well as with experiment. Furthermore, owing to the full rotational
computations performed, the accuracy of the resulting dissociative recombination
cross sections is considerably improved.
\end{abstract}


\maketitle

\section{\textbf{Introduction }}

In the modeling of the kinetics of cold dilute gases,
the rotational distribution of molecular species is governed by competition
between formation and destruction processes, absorption, fluorescence, radiative
cascades and low-energy
collisions involving neutral and ionized atomic and molecular species
as well as electrons.
Rate coefficients for such elementary reactions are badly needed, in particular
for the chemical
models of the early Universe,
interstellar media and planetary atmospheres \cite{coppola,roueff,black,gay}.

Much has been done during the past years to improve the description of
rotational transitions induced in molecular cations by collision with
atoms or molecules. Most of the computations are based on the close
coupling technique implemented in computer codes like MOLSCAT
\cite{molscat}. Meanwhile, in diffuse environments, electrons are
expected to be the dominant exciting species for molecular ions as the
cross sections for electron impact excitation are several orders of
magnitude greater than the corresponding ones for excitation by
neutral atomic or molecular species. Computations for electron induced
rotational excitation and de-excitation have been previously performed
for diatomic and linear triatomic molecular ions (CH$^+$, HeH$^{+}$,
NO$^{+}$, H$_{2}^{+}$, CO$^{+}$, HCO$^{+}$)
\cite{lim99,rabadan98b,faure01,JS06}, as well as polyatomic molecular
ions (H$_{3}^{+}$, D$_{3}^{+}$, H$_{3}$O$^{+}$, D$_{3}$O$^{+}$, etc.)
\cite{faure02,faure03,faure06,kokoo10}. The theory is based on the
R-matrix method \cite{jt474} augmented by use of the Coulomb-Born
approximation to account for long-range dipole interactions, and the
adiabatic-nuclei rotation (ANR) approximation
\cite{faure02,rabadan98a}. This method avoids consideration of an
excessively large number of channels in a rotational close-coupling
expansion.  In the specific case of the near-threshold rotational
excitation of H$_{3}^{+}$ by electron impact, the cross sections
obtained using the ANR/R-matrix method have been shown to be in good
agreement with those coming from Multichannel Quantum Defect Theory
and rotation frame transformation (MQDT-RFT) \cite{faure06}.

The hydrogen molecular ion, which is thought to participate in the chemistry of
harsh environments and considered as a key species in the formation of H$_{2}$
in the early universe \cite{rawlings,coppola} is an important subject of
investigation. Its dissociative recombination (DR) together with the competing
reactions - rovibrational excitation/de-excitation and dissociative
excitation - play a decisive role in astrophysical ionized media
(stars and interstellar molecular clouds)\cite{coppola,roueff,black}, fusion plasma in the
divertor region \cite{janev,reiter,fantz} and in most of the hydrogen-containing cold
plasmas of technological interest. Recent advanced
studies on H$_{2}^{+}/e^{-}$
\cite{deruette07,zhaunerchyk07,mot08} have resulted in very accurate state-to-state
rate-coefficients. However, the HD$^{+}/e^{-}$
deuterated version of the H$_{2}^{+}/e^{-}$ benchmark system has
received more attention in  storage-ring-type
experiments, since its permanent dipole facilitates  rapid rotational and
vibrational cooling of the ion. Consequently, the dissociative
recombination (DR) of  ${\text H}{\text D}^{+}$:

\begin{equation}
\label{eq:DR} {\text H}{\text D}^{+} + e^{-} \longrightarrow {\text
H} + {\text D}^{*}, {\text H}^{*} + {\text D}
\end{equation}

\noindent has been extensively and systematically studied both experimentally and theoretically
\cite{zajfman93,takagi95,stromholm95,andersen95, ifs-a18,amitay98,
amitay99, takagi03, alkhalili04,buhr06,waffeu11}.

More recently, the rotational cooling of ${\text H}{\text D}^{+}$ molecular ions by
superelastic collisions:

\begin{equation}
\label{eq:SEC} {\text H}{\text D}^{+}(v_{i}^{+}=0, N_{i}^{+}) +
e^{-}(\varepsilon) \longrightarrow
{\text H}{\text D}^{+}(v_{i}^{+}=0, N_{f}^{+}) +
e^{-}({\varepsilon}^{'})
\end{equation}

\noindent where $N_{f}^{+} < N_{i}^{+}$, was investigated at the Test
Storage Ring (TSR) of the MPIK at Heidelberg \cite{shafir,schwalm}.
By merging an ${\text H}{\text D}^{+}$ beam with velocity matched
electrons, rapid cooling of the rotational excitations of the ${\text
  H}{\text D}^{+}$ ions by superelastic collisions (SEC) with the
electrons was observed. The cooling process, which was monitored by
the time evolution of the relative populations of groups of rotational
levels, was well described using theoretical SEC rate coefficients
obtained by combining the molecular R-matrix approach with the
adiabatic nuclei rotation approximation.  The present work is aimed at
computing cross sections and rate coefficients for state-to-state
rotational transitions (inelastic collisions and superelastic
collisions) within the framework of the stepwise-method based on the
Multichannel Quantum Defect Theory \cite{annick80,mot08}, and then
providing a comparison with the recent experiment as well as the
adiabatic nuclei rotation approximation. Such data may constitute a
valuable basis for a critical interpretation of the observed spectra
of the species of interest \cite{coppola,gay,roueff,black}. Moreover,
the results are obtained through a series of full rotational
computations similar to the treatment of Ref.~\cite{mot08} for
H$_{2}^{+}$, where all the relevant symmetries were appropriately
considered. This provides us with improved ${\text H}{\text D}^{+}$
DR rate coefficients in comparison to our
recent work \cite{waffeu11} where the computations were restricted to
the dominant $^1\Sigma_{g}^{+}$ symmetry.

This paper is organized as follows:
the main steps in the computation of the cross section,
based on our MQDT-type method, are reviewed in Section II, the computation of
cross sections and rate coefficients and detailed comparison with former
results are described in section III, and the conclusions are provided in section IV.

\section{\textbf{Theoretical Method}} 
The reactive collisions between electrons and molecular cations
involve \emph{ionization} channels, describing the scattering of an electron
on the molecular ion and
\emph{dissociation} channels, accounting for atom-atom scattering. The main
steps of our current MQDT treatment \cite{mot08} are described below:

\subsection{Construction of the interaction matrix \Vmat}

Construction of the interaction matrix, \Vmat, is performed in the outer shell of
the region of small
electron-ion and nucleus-nucleus distances, that is, in the `A-region'
\cite{ja77} where the Born-Oppenheimer approximation gives an 
appropriate description of the collision system. The
good
quantum numbers in this region are $N$, $M$, and $\Lambda$, associated respectively
with the total
angular momentum  and its projections on the
z-axis of the laboratory-fixed and of the molecule-fixed frame.

Within a \emph{quasi-diabatic representation} \cite{bardsley68,sidis72,annick80}, the relevant states are organized in \emph{channels}, according
to the type of fragmentation which they are meant to describe.
An \emph{ionization} channel
is built starting from the ground electronic state
of the ion and one of its ro-vibrational levels $N^{+} v^{+}$, and is completed by gathering all the mono-electronic states
of a given orbital quantum number $l$. These mono-electronic states describe, \emph{with respect to the
$N^{+}v^{+}$ threshold}, either a "\emph{free}" electron - in which case the total state
  corresponds to \emph{(auto)ionization} -  or to a \emph{bound} electron - in which case, the total state
  corresponds to a temporary \emph{capture into a Rydberg state}. In the A-region, these states
may be modeled reasonably well with respect to the hydrogenic states in terms of
 the quantum defect $\mu^{\Lambda}_{l}$, dependent on the internuclear distance
 $R$,  but assumed to be \emph{independent of energy}.

An ionization channel is coupled to a
\emph{dissociation} one, labeled
$d_{j}$, by the electrostatic interaction $1/r_{12}$. In the molecular-orbital picture,
the states corresponding to the coupled channels must differ by at least two
orbitals, the dissociative states being doubly-excited
for the present case. We account for this coupling at the \emph{electronic} level
first, through an $R$-dependent \emph{scaled} "Rydberg-valence" interaction
term,  $V^{(e)\Lambda}_{d_{j},l}$, assumed to be
\emph{independent of the energy} of the electronic states pertaining to the
ionization channel. Subsequently, the integration of this electronic
interaction on the internuclear motion results in the elements of the
interaction matrix \Vmat:

\begin{equation}
\label{vmat}
 \mathcal{V}_{d_{j},lN^{+}v^{+}}^{NM\Lambda}(E,E)=
\langle\chi^{N\Lambda}_{d_{j}}|V^{(e)\Lambda}_{d_{j},l}|\chi^{\Lambda}_{N^{+},v^{+}}\rangle
\end{equation}
where $E$ is the total energy and
$\chi^{\Lambda}_{Nd_{j}}$ and $\chi^{\Lambda}_{N^{+},v^{+}}$ are the nuclear
wave-functions corresponding to a
dissociative state and to an ionization channel respectively.

This procedure applies in each $\Lambda$-subspace and results in a block-diagonal
global interaction matrix. The block-diagonal structure, corresponding to the 
$\Lambda$ symmetries, propagates to the other matrices computed.

\subsection{Construction of the reaction matrix \Kmat}
Starting from the interaction matrix \Vmat, we build the
\Kmat-matrix, which satisfies the Lippmann-Schwinger integral
equation:
{\boldmath
\begin{equation}
\label{eq:Lippmann-Schwinger}{\cal K}={\cal V} + {\cal V}
\frac{1}{E-H_0} {\cal K}
\end{equation}
}

\noindent
This equation has to be solved once \Vmat\quad- whose elements are given by eq. (\ref{vmat}) - is determined.
Here {\boldmath $H_0$} is the zero-order Hamiltonian
associated with the molecular system neglecting the interaction potential \Vmat.
It has been proven that, provided the electronic couplings are energy-independent,  a perturbative solution of
equation~(\ref{eq:Lippmann-Schwinger}) is exact to the second order
\cite{ngassam03b}.

\subsection{Diagonalization of the reaction matrix \Kmat}
In order to express the result of the short-range interaction in terms of phase-shifts,
we perform a unitary transformation of our initial basis into eigenstates.
The columns of the corresponding transformation unitary matrix {\Umat}
are the eigenvectors of the K-matrix:

\begin{equation}
{\cal K} {\mathbf \it U}= -{\frac{1}{\pi}}\tan(\eta){\mathbf \it U}
\label{K-pvp}
\end{equation}

\noindent
and its eigenvalues, expressed as
the non-vanishing elements of a diagonal matrix
$-{\frac{1}{\pi}}\tan(\eta)$, provide the diagonal matrix of the phaseshifts
 $\eta$, induced in the eigenstates by the short-range interactions.

\subsection{Frame transformation to the external region}

In the external zone - the `B-region' \cite{ja77} - characterized by
large electron-core distances, the Born-Oppenheimer representation
is no longer valid for the whole molecule, but only for the ionic core. Here
$\Lambda$ is no longer a good quantum number
and a frame
transformation \cite{fano70,chfan,valcu} is performed between coupling
schemes corresponding to the incident electron being decoupled
from the core electrons
(external region) or coupled to them (internal region).
The frame transformation
coefficients involve angular coupling
coefficients, electronic and ro-vibronic factors, and are given by:

\begin{equation}
\label{eq:coeffCv}
\begin{array}{l}
{\cal C}_{lN^{+}v^{+}, \Lambda \alpha}=\left(
\frac{2N^{+}+1}{2N+1}\right) ^{1/2}\left\langle l\left( \Lambda
-\Lambda ^{+}\right) N^{+}\Lambda ^{+}|lN^{+}N\Lambda
\right\rangle \\ \begin{array}{c}
 \times\frac{1+\tau^{+}\tau\left(-1 \right)^{N-l-N^{+}}} {\left[2\left(2-\delta_{\Lambda^{+},0} \right)\left(1+ \delta_{\Lambda^{+},0}\delta_{\Lambda,0} \right)   \right] ^{1/2}}\\ \begin{array}{c}

\times\sum_{v} U_{lv,\alpha}^{\Lambda}\langle
\chi_{N^{+}v^{+}}^{\Lambda ^{+}}| \cos(\pi \mu_{l}^{\Lambda}
(R)+\eta_{\alpha}^{\Lambda})|\chi_{Nv}^{\Lambda}\rangle
\end{array}\end{array}\end{array}
\end{equation}

\begin{equation}
\label{eq:coeffCd}{\cal C}_{d_{j},\Lambda
\alpha}=U_{d_{j}\alpha}^{\Lambda}\cos \eta_{\alpha}^{\Lambda}
\end{equation}

\noindent as well as ${\cal S}_{lN^{+}v^{+},\Lambda \alpha }$ and
${\cal S}_{d_{j},\Lambda \alpha }$, which are obtained  by replacing cosine
with sine in Eqs. (\ref{eq:coeffCv}) and (\ref{eq:coeffCd}).
In the preceding formulas, $\chi_{N^{+}v^{+}}^{\Lambda ^{+}}$ is a
vibrational wavefunction of the molecular ion, and
$\chi_{Nv}^\Lambda$ is a vibrational wavefunction of the neutral
system adapted to the interaction (A) region. The quantities
$\tau^{+}$ and $\tau$
  are related to the reflection symmetry of the ion and neutral wave
  function respectively, and take the values +1 for symmetric states and -1 for
  antisymmetric ones. The ratio in front of the sum in the right hand side of
  Eq. (\ref{eq:coeffCv}) contains the selection rules for the rotational quantum numbers.
The indices $d_{j}$ ($j=0, 1, 2,...$) stand for the states of a
given symmetry that are open to dissociation at the current energy.
Here $\alpha$ denotes the eigenchannels built through the {\it
diagonalization} of the reaction matrix
 \Kmat\, and $-\tan(\eta_{\alpha}^{\Lambda})/\pi$, $U_{lv,\alpha}^{\Lambda}$ are its eigenvalues and the components of its eigenvectors respectively.

The projection coefficients shown in Eqs. (\ref{eq:coeffCv}) and
(\ref{eq:coeffCd}) include the two types of couplings that control
the process: the {\it electronic} coupling, expressed by the
elements of the matrices \Umat\ and \etamat, and the {\it
non-adiabatic} coupling between the ionization channels, expressed
by the matrix elements involving the quantum defect
$\mu_{l}^{\Lambda}$. This latter interaction is favored by the
variation of the quantum defect with the internuclear distance
$R$.

\subsection{Construction of the generalized matrix \Xmat }

The matrices \Cmat\  and  \Smat\  with the elements given by Eqs.
(\ref{eq:coeffCv}) and (\ref{eq:coeffCd}) are the building blocks of the `generalized' scattering matrix \Xmat:
\begin{equation}
\label{eq:Xmatrix}\Xmat =\frac{\Cmat+i\Smat}{\Cmat-i\Smat}
\end{equation}

\noindent
It involves all the channels, open 
and closed.
Although
technically speaking only the open channels are relevant for a complete
collision event, the participation of the closed channels may influence
strongly the cross section, as  shown below.

The {\Xmat} matrix relies on 4 block sub-matrices:
\begin{equation}
\label{eq:blocks}
\Xmat =
\left(
  \begin{array}{cc}
    X_{oo} & X_{oc} \\
    X_{co} & X_{cc} \\
  \end{array}
\right)
\end{equation}
\noindent
where $o$ and $c$ label the lines or columns corresponding to \emph{open} and \emph{closed} channels, respectively.

subsection{Elimination of closed channels}

The building of the X matrix is performed independently of any account of the asymptotic behavior of the different channel wavefunctions.
Eventually, imposing physical boundary conditions leads to the `physical' scattering matrix, restricted to
the {\it open} channels~\cite{seaton83}:
\begin{equation}
\label{eq:elimination}\SSmat=\Xmat_{oo}-\Xmat_{oc}\frac{1}{\Xmat_{cc}-\exp({\rm
-i 2 \pi} \numat)} \Xmat_{co}\end{equation} \noindent
This scattering matrix is
obtained from the sub-matrices of \Xmat\
appearing in Eq. (\ref{eq:blocks})
and from a further diagonal matrix \numat\  formed with the effective
quantum numbers ${\nu}_{N^{+}v^{+}}=[2(E_{N^{+}v^{+}}-E)]^{-1/2}$
(in atomic units) associated with each vibrational threshold
$E_{N^{+}v^{+}}$ of the ion situated {\it above} the current
energy $E$ (and consequently labelling a \emph{closed} channel).

\subsection{Cross sections evaluation}

For a molecular ion initially on the level $N_{i}^{+}v_{i}^{+}$
and recombining with an electron of kinetic energy $\varepsilon$, the
cross section of capture into {\it all} the dissociative states
$d_{j}$ of the same symmetry is given by:
\begin{equation}
\label{eq:cs-partial}\sigma _{diss \longleftarrow
N_{i}^{+}v_{i}^{+}}^{N,sym}=\frac{\pi }{4\varepsilon
}\frac{2N+1}{2N_{i}^{+}+1}\rho^{sym}\sum_{l,\Lambda,j}
 |S^{{N\Lambda}}_{d_{j},l N_{i}^{+}v^{+}_{i}}|^{2}
\end{equation}

On the other hand, the cross section for a rovibrational transition to the
final level $N_{f}^{+}v_{f}^{+}$, giving collisional (de-) excitation is:

\begin{equation}
\label{eq:partcs} \sigma _{N_{f}^{+}v_{f}^{+} \longleftarrow
N_{i}^{+}v_{i}^{+}}^{N,sym}=\frac{\pi }{4\varepsilon
}\frac{2N+1}{2N_{i}^{+}+1}\rho^{sym}\sum_{l,l',\Lambda,j}
\left\vert
S_{N_{f}^{+}v_{f}^{+}l',N_{i}^{+}v_{i}^{+}l}^{N\Lambda}\right\vert
^{2}
\end{equation}
\noindent Here $\rho ^{sym}$ is the ratio between the
multiplicities of the neutral and the target ion. After performing
the MQDT calculation for all the accessible total rotational
quantum numbers $N$ and for all the relevant  symmetries, one has
to add up the corresponding cross sections in order to obtain the
global cross section for dissociative recombination or
rovibrational transition, as a function of the electron collision energy $\varepsilon$.

\section{\textbf{Computation of cross sections and rate coefficients}}

\subsection{\textbf{Molecular data}}

Although the lowest  $^1\Sigma_{g}^{+}$ doubly excited state has
the most favorable crossing with the ion curve for collisions taking place at low electronic
energy, a comprehensive analysis requires the consideration of the contribution of the other
molecular symmetries to  the cross
section, especially as rotational couplings have been found to enhance
superelastic and inelastic collisions in this energy range. Therefore, the
molecular states that are included are $^1\Sigma_{g}^{+}$, $^1\Pi_{g}$,
$^1\Delta_{g}$, $^3\Sigma_{g}^{+}$, $^3\Pi_{g}$,
$^3\Delta_{g}$, $^3\Sigma_{u}^{+}$, and $^3\Pi_{u}$. The $^1\Sigma_{g}^{+}$
symmetry is treated simultaneously with the symmetric components with respect to
the reflection symmetry of $^1\Pi_{g}$ and $^1\Delta_{g}$ to account for
rotational couplings between them. A similar treatment is used for the triplet
gerade symmetries, and for $^3\Sigma_{u}^{+}$ and $^3\Pi_{u}$.
The antisymmetric components with respect to
the reflection symmetry of $^{1,3}\Pi_{g}$ and $^{1,3}\Delta_{g}$ are neglected as
well as the $^1\Sigma_{u}^{+}$ and $^1\Pi_{u}$, because their contribution
to the total cross sections is very small in comparison to the other ones.

For the $^{1}\Sigma _{g}^{+}$\ state, we have used the adjusted
quasi-diabatic molecular data - electronic potential curves and couplings -
previously used in Refs. \cite{mot08} and \cite{waffeu11}. The data for $^{3}\Sigma _{g}^{+}$, $^{1}\Delta _{g}$%
\ and $^{3}\Delta _{g}$ have been obtained by combining the data of Tennyson
\cite{tennyson96} with results of new computations using the halfium code developped by
Telmini and Jungen \cite{TJ}. In this code, the variational eigenchannel
R-matrix method and the generalized
multichannel quantum defect theory are combined in prolate spheroidal
coordinates to determine the potential electronic energy curves of H$_{2}$.
The configuration space is divided into two zones: a reaction zone and an
external zone. In the inner zone, the Schr\"{o}dinger equation is solved
using the variational R-matrix method. The variational basis takes into
account spin
and $\sigma _{v}$ symmetrizations allowing for the description of all
molecular symmetries including $\Sigma ^{-}$ symetries \cite{BATJ,S-ABOTJ}.
In the external zone, the hydrogen molecule is modeled as a
three-body system; two positive half-charge nuclei and one electron. This
approximation, called the halfium model, captures the partial screening of the
protons by the internal electron \cite{BT}. Matching the two expressions of
the wavefunctions in the two zones leads to the reaction matrix which
contains the short-range interaction between the different possible channels.

Both Rydberg and doubly excited states are computed on the same footing.
Thanks to the generalized MQDT formalism, all members of any Rydberg series are easily
computed for arbitrary high principal quantum number. Therefore, the quantum
defects of the highest Rydberg states up to the ionisation threshold are
deduced in a straightforward procedure. Positions and widths of doubly
excited states are obtained from resonance Breit-Wigner profiles of state
density~\cite{tbj}, and couplings are deduced from widths.

Previous results obtained for lowest Rydberg states \cite{TJ,BTJ} 
agree well with the available {\em ab initio} data \cite{wol3}. Energies and
widths of doubly excited states \cite{TJ,BATJ,tbj,BTJ}
) are also in good agreement with those of Tennyson \cite{tennyson96}. \ In this work we
extended the calculations towards higher principal quantum number and larger
internuclear distances as well. As quasi-diabatic potentiel energy curves
are needed, the closed channels are treated as open through the introduction
of an artificial threshold. In this way, the data for doubly-excited states
including energies and couplings beyond the crossing point between these
states and the molecular ion's ground state electronic curve are obtained.
Dissociation limits are deduced from the correlation diagrams \cite{lim diss}
and extrapolation of potential energy curves is
performed until the energy corresponding to the dissociation products is reached.

For each of the symmetries involved -
$^{1,3}\Sigma_{g}^{+}$,
$^{1,3}\Pi_{g}$,
$^3\Sigma_{u}^{+}$,
$^3\Pi_{u}$,
$^{1,3}\Delta_{g}$ -
only the lowest dissociative state, relevant for
low energy computation, is
considered.

Two partial waves ($s$ and $d$) have been taken into account for the
$^1\Sigma_{g}^{+}$
state and only one partial wave for the others ($d$ for $^1\Pi_{g}$,
$^1\Delta_{g}$, $^3\Sigma_{g}^{+}$, $^3\Pi_{g}$ and
$^3\Delta_{g}$, and $p$ for $^3\Sigma_{u}^{+}$, and $^3\Pi_{u}$). The
symmetries $^1\Delta_{g}$ and $^3\Delta_{g}$ do not have an open
dissociative channel in the energy range of
interest, but  they contribute through the indirect mechanism via
their coupling to the rotationally compatible symmetries.
In practice this is also the case for the $^3\Sigma_{g}^{+}$ symmetry, whose lowest dissociative
state is open but is
very weakly coupled to the ionization continuum.

\subsection{\textbf{Results and comparison with experiment}}

Relying on the data described above,
we have performed MQDT-based
\cite{annick80,mot08} full rotational computations of the cross sections of 
electron induced rotational transitions 
and 
dissociative recombination 
for
vibrationally relaxed HD$^{+}$ molecular ions on their first 11 rotational levels
($v_{i}^{+}=0, N_{i}^{+}=0-10$), which are easily sufficient to
satisfactorily model an equilibrium rotational distribution
corresponding to temperatures below 1000 K. As mentioned above, rotationally compatible symmetries
are treated simultaneously on the same footing.

For each initial rotational state, $N_{i}^{+}$, for the
ion in its
ground vibrational level, $v_{i}^{+}=0$, the cross section for the rotational transition
$\sigma _{N_{f}^{+}v_{i}^{+} \longleftarrow N_{i}^{+}v_{i}^{+}}$ is obtained from its
partial contributions $\sigma _{N_{f}^{+}v_{i}^{+} \longleftarrow N_{i}^{+}v_{i}^{+}}^{N,sym}$
by summing them up over all the
accessible values of $N$ and all the symmetries. In a similar way, the DR cross section
 $\sigma_{diss\longleftarrow N_{i}^{+}}$ is obtained from its
 partial contributions
 $\sigma_{diss\longleftarrow N_{i}^{+}}^{N,sym}$.

\begin{figure}
\begin{center}
\includegraphics[width=0.95\columnwidth]{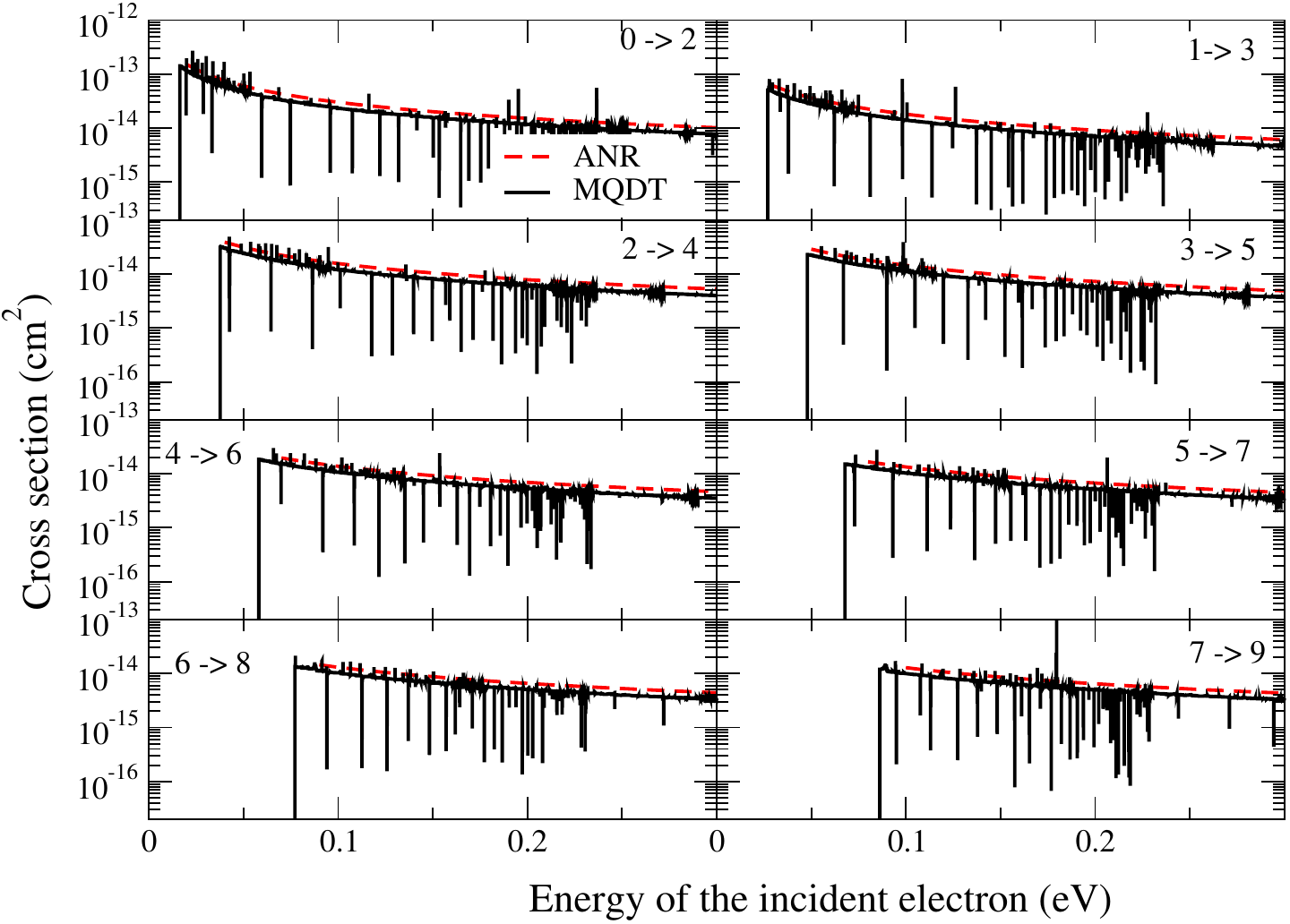}
\end{center}
\caption{\label{cap:cs_all} (Color online){\textit{
Cross sections  for rotational excitations
$N_{i}^{+}$-2 $\rightarrow$ $N_{i}^{+}$, with $N_{i}^{+}$=2 to 9, in the ground
vibrational state of HD$^{+}(X^{2}\Sigma_{g}^{+})$. Black solid curves: MQDT computations;
red dashed 
curves: ANR approximation based
computations.}}}
\end{figure}

\begin{figure}
\begin{center}
\includegraphics[width=0.95\columnwidth]{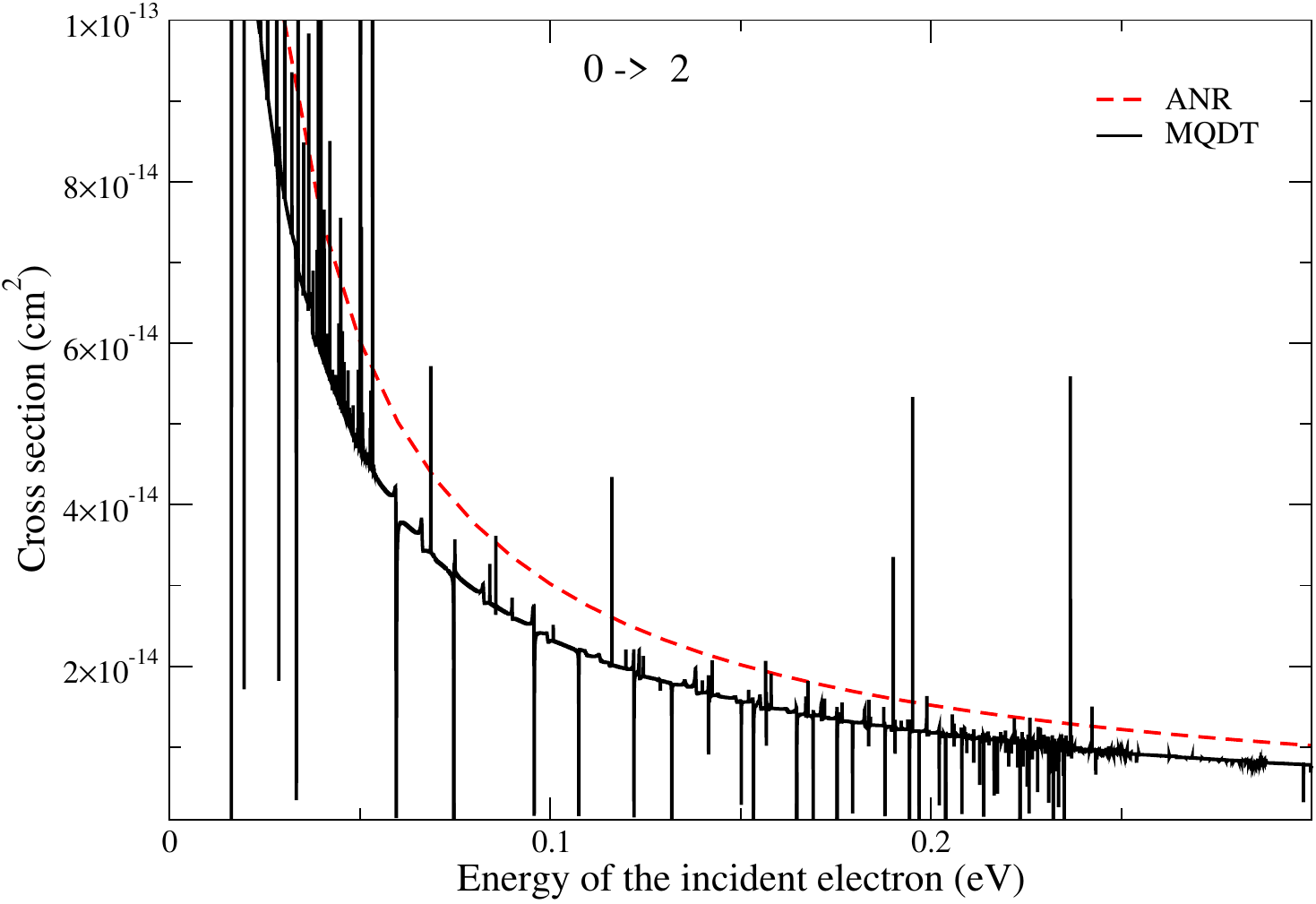}
\end{center}
\caption{\label{cap:cs_02b}(Color online) {\textit{Zoom in the linear scale of the cross section  for the transition
0 $\rightarrow$ 2,  in the ground
vibrational state of HD$^{+}(X^{2}\Sigma_{g}^{+})$. Black solid curve: MQDT computations;
red dashed curve: ANR approximation based
computations. }}}
\end{figure}

The cross sections for rotational excitations ($N_{i}^{+}$ $\rightarrow$
$N_{i}^{+}+2$) are represented in Figure \ref{cap:cs_all}, where they are
compared with ones computated using the R-matrix/ANR method, see Ref.~\cite{shafir} for details. A zoom of the cross
section for the transition 0$\rightarrow$2 is presented in Figure
\ref{cap:cs_02b}. The MQDT results exhibit a rich resonance structure which is
due to the indirect process that is the temporary capture into the numerous
Rydberg states of the neutral system (HD) involved in the process. The cross
sections from the ANR approximation do not present such resonances
since this approach neglects the role of the indirect process.
The competition with the
DR is also neglected within the ANR approach.
However, in spite of the relative simplicity of this method,
an overall good
agreement is found between the two approaches.
Nevertheless, it should be noted that the
 MQDT computations treat
the HD molecule  like H$_2$ from the electronic point of view.
Consequently, the {\it gerade} symmetry is completely uncoupled from the {\it ungerade}
one and, due to selection rules inherent to the relevant partial waves
(see Eq. (\ref{eq:coeffCv})), only transitions with
even $\Delta N^{+}$
(0, 2 and 4) are allowed, 
whereas  the ANR/R-matrix computations considers transitions with
odd $\Delta N^{+}$ (1 and 3), whose rate coefficients are not negligible but are smaller by a factor of 2 to 4.

\begin{figure}
\begin{center}
\includegraphics[width=0.95\columnwidth]{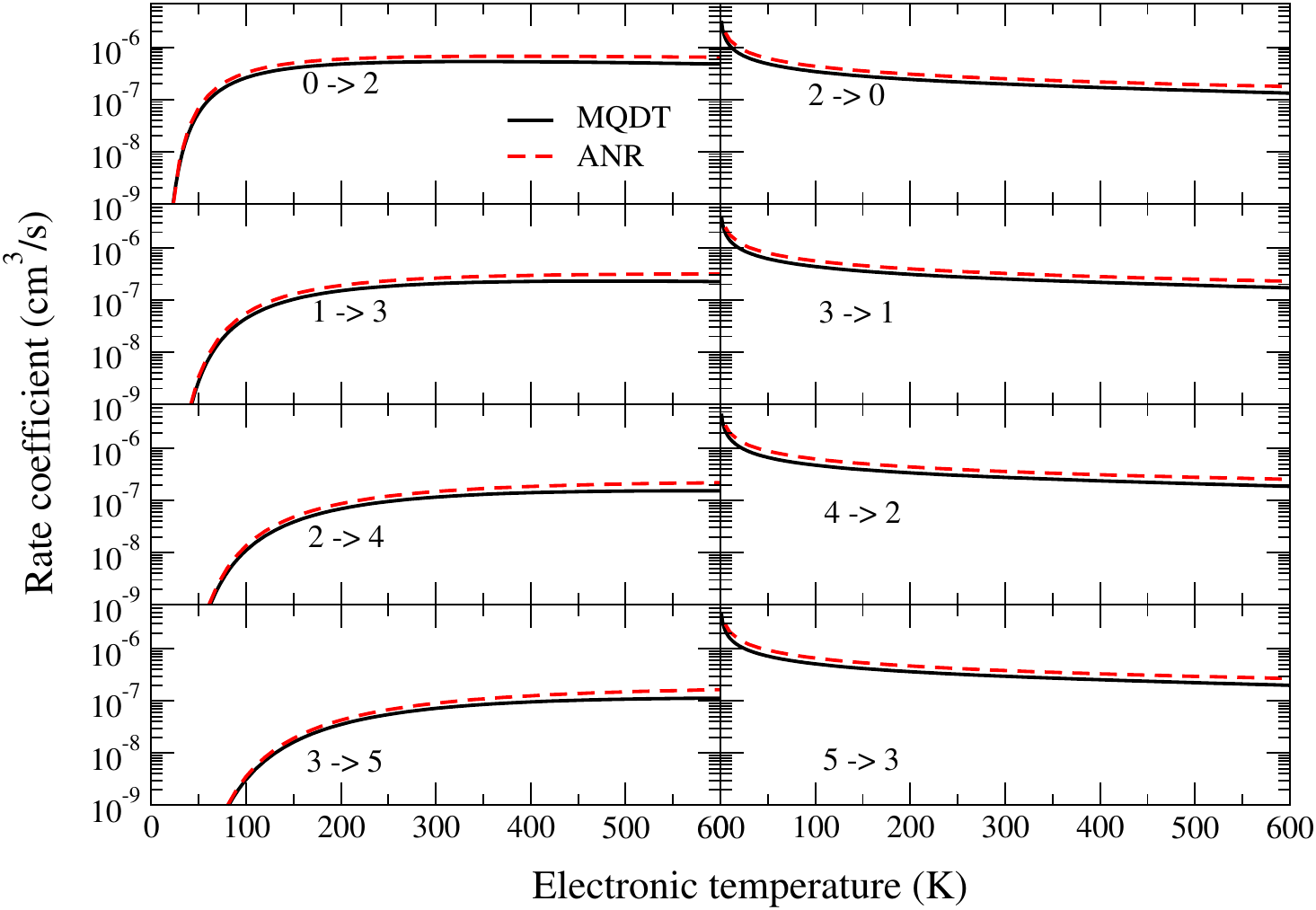}
\end{center}
\caption{\label{cap:r_all1} (Color online){\textit{Rate coefficients of the rotational
excitations $N_{i}^{+}$-2 $\rightarrow$ $N_{i}^{+}$, with $N_{i}^{+}$=2 to 5, and reverse reactions, for the ground
vibrational state of HD$^{+}(X^{2}\Sigma_{g}^{+})$. Black solid curves: MQDT computations;
red dashed curves: ANR approximation
based computations.}} }
\end{figure}

\begin{figure}
\begin{center}
\includegraphics[width=0.95\columnwidth]{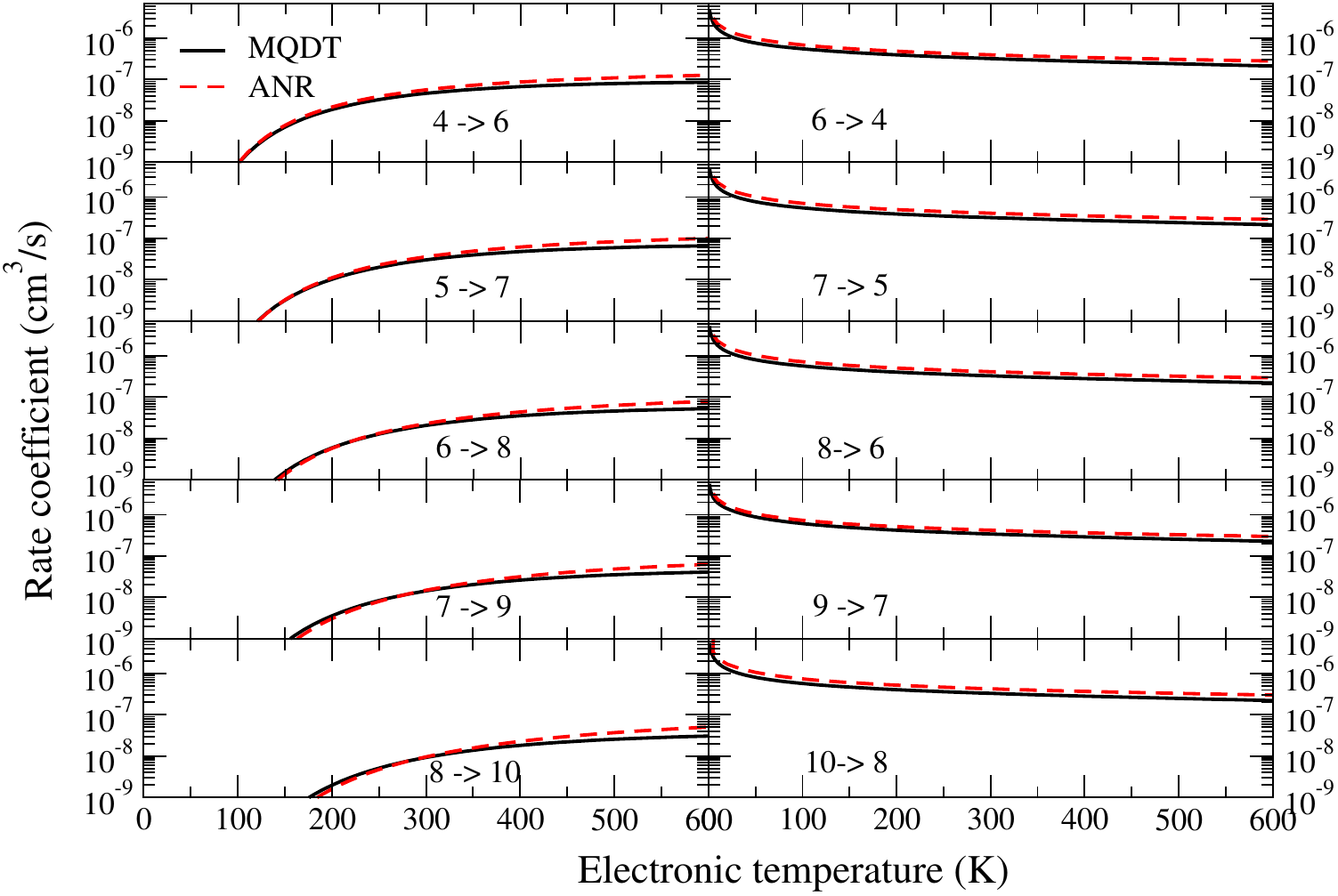}
\end{center}
\caption{\label{cap:r_all2} (Color online){\textit{Rate coefficients of the rotational
excitations $N_{i}^{+}$-2 $\rightarrow$ $N_{i}^{+}$, with $N_{i}^{+}$=6 to 10, and reverse reactions for the ground
vibrational state of HD$^{+}(X^{2}\Sigma_{g}^{+})$. Black solid curves: MQDT computations;
red dashed  curves: ANR approximation based
computations.}} }
\end{figure}

Isotropic and anisotropic Maxwell rate coefficients were
obtained for the rotational transitions by appropriate convolution of 
the computed cross-sections. Good  agreement can  be
observed in Figs.~\ref{cap:r_all1} and \ref{cap:r_all2} where Maxwell isotropic
rate coefficients are compared for the transitions of interest (excitation and
de-excitation).

The {\emph{anisotropic}} Maxwell rate coefficients at near-zero
energy, obtained by convolution of the cross sections with the
distribution of the relative electron-ion velocity in the latest
SEC-dedicated TSR storage ring experiments \cite{shafir,schwalm}
(characterized by transversal and longitudinal temperatures
$k_{B}T_{t}=2.5$meV and $k_{B}T_{l}$=45$\mu$eV respectively) are
presented in Table \ref{sectab}, and comparison with the ANR/R-matrix calculations is performed.
The values displayed are close to the average experimental rate coefficient of 1.7$\cdot$10$^{-6}$ 
cm$^{3}/$s found by Schwalm et al.  \cite{schwalm}. The good agreement between the computed values and the measured ones confirms the dominant role of the superelastic collisions
on the rotational cooling of HD$^{+}$.

\begin{table}
\caption{Anisotropic rate coefficients for superelastic collisions of
HD$^{+}$ with electrons of near-zero kinetic energy (in units of $10^{-6}$
cm$^3$/s).}
\label{sectab}%
\begin{ruledtabular}
\begin{tabular}{ccc}
 Transition & MQDT (this work) & R-matrix/ANR  \cite{shafir,schwalm} \\
\hline
$10 \rightarrow 8$  &  1.46 &  1.82  \\
 $9 \rightarrow 7$  &  1.57 &  1.81  \\
 $8 \rightarrow 6$  &  1.44 &  1.79  \\
 $7 \rightarrow 5$  &  1.39 &  1.76  \\
 $6 \rightarrow 4$  &  1.36 &  1.72  \\
 $5 \rightarrow 3$  &  1.28 &  1.00  \\
 $4 \rightarrow 2$  &  1.22 &  1.19  \\
 $3 \rightarrow 1$  &  1.11 &  1.19  \\
 $2 \rightarrow 0$  &  0.87 &  0.98  \\

\end{tabular}
\end{ruledtabular}
\end{table}

\begin{figure}
\begin{center}
\includegraphics[width=0.95\columnwidth]{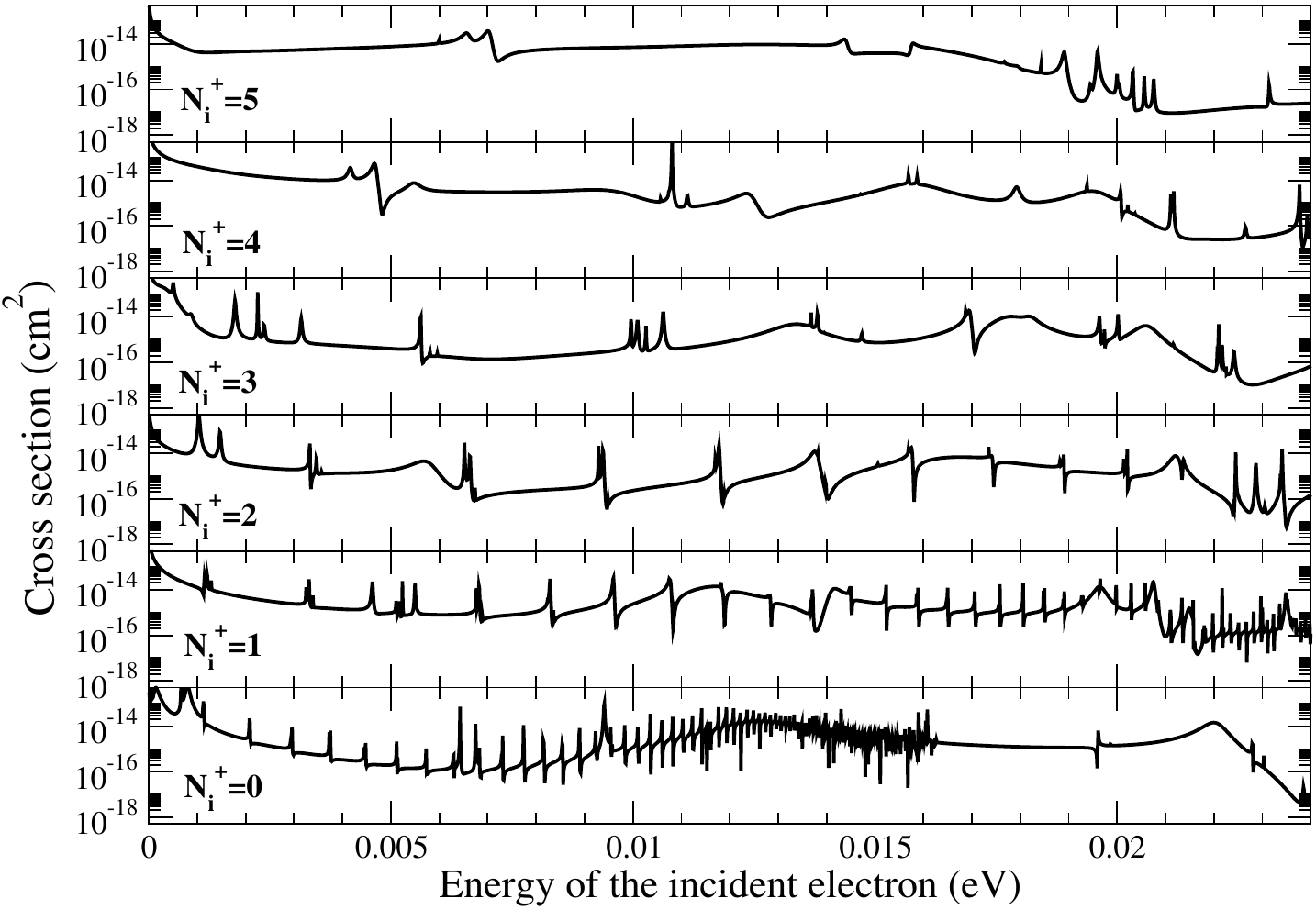}
\end{center}
\caption{\label{cs_DR_1}
{\textit{Cross sections for the
Dissociative Recombination
of vibrationally relaxed
HD$^{+}(X^{2}\Sigma_{g}^{+})$
on initial rotational levels $N_{i}^{+}$=0 to 5
for electron collision energy
in the range 0-24 meV}.}}
\end{figure}

\begin{figure}[bt]
\begin{center}
\includegraphics[width=0.95\columnwidth]{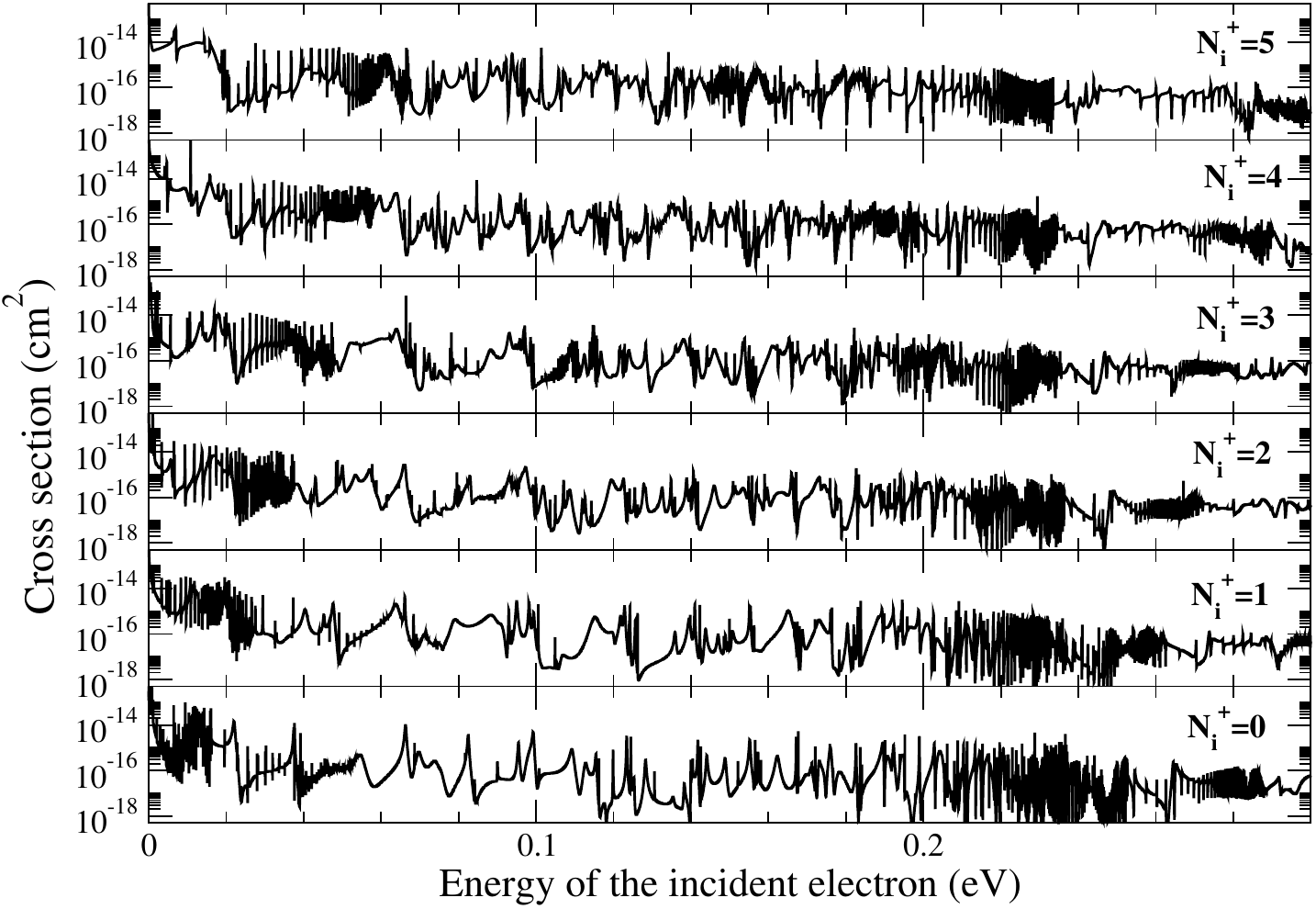}
\end{center}
\caption{\label{cs_DR_2}
{\textit{Cross sections for the
Dissociative Recombination
of
vibrationally relaxed
HD$^{+}(X^{2}\Sigma_{g}^{+})$
on
initial rotational levels $N_{i}^{+}$=0 to 5
for
electron collision energies
in the range 0-300 meV}.}}

\end{figure}

\begin{figure}
\begin{center}
\includegraphics[width=0.95\columnwidth]{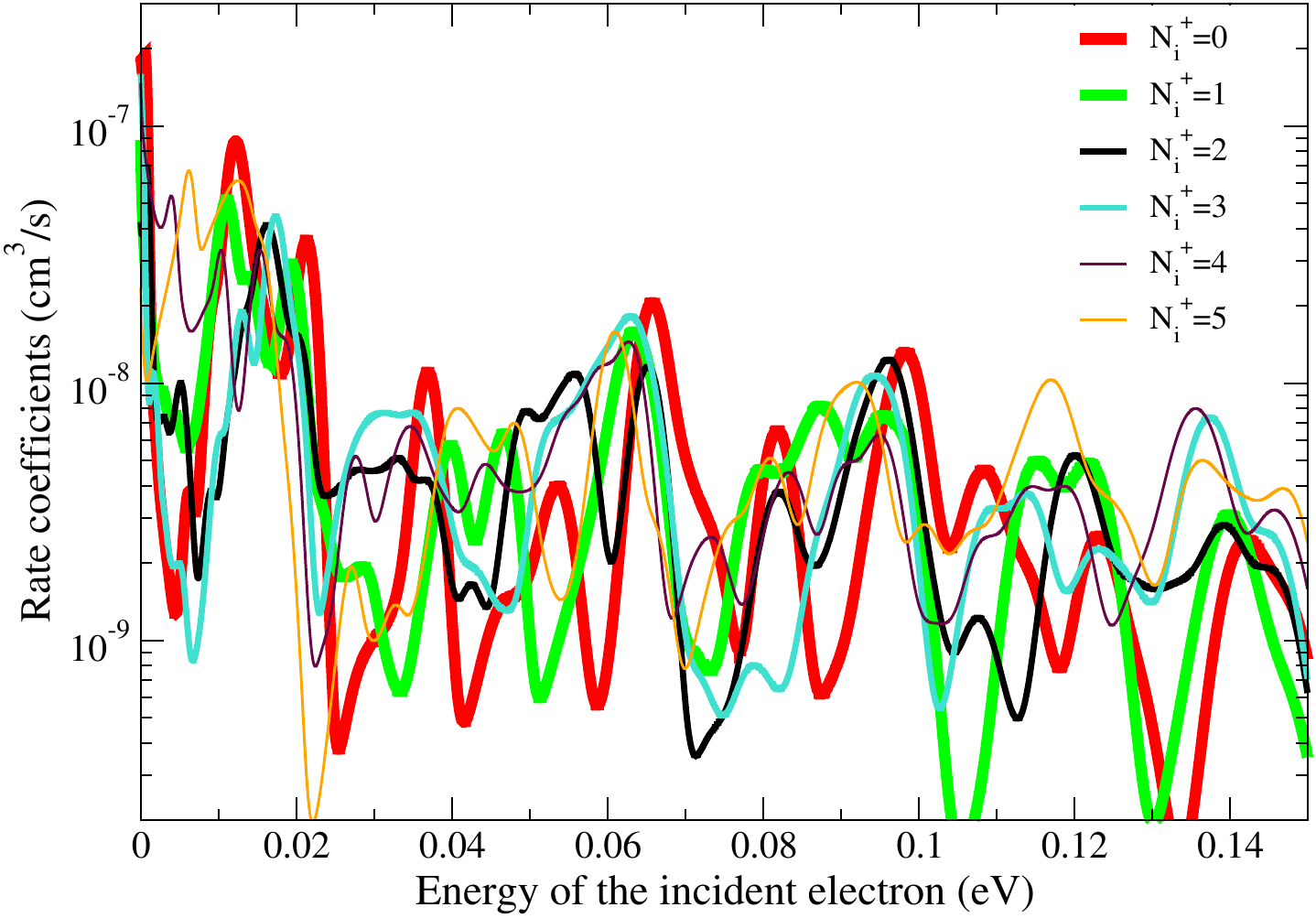}
\end{center}
\caption{\label{rate_ani_DR_2}(Color online) {\textit{Maxwell anisotropic rate coefficients for the
Dissociative Recombination
of vibrational relaxed
HD$^{+}(X^{2}\Sigma_{g}^{+})$
on its
initial rotational levels $N_{i}^{+}$=0 to 5
as in \cite{alkhalili04, buhr06,waffeu11}}.}}
\end{figure}

Within the same MQDT computations, new DR
cross sections are obtained for all the 
rotational states of vibrationally relaxed HD$^{+}(X^{2}\Sigma_{g}^{+})$ of
interest in this work. 
Those for the first six initial rotational levels ($N_{i}^{+}$=0 to 5) are plotted in
Figures \ref{cs_DR_1} and \ref{cs_DR_2} for the energy ranges up to 24 meV  and  300
meV respectively.
From the DR cross sections obtained, Maxwell anisotropic rate coefficients in the conditions of the latest DR-dedicated TSR experiments \cite{waffeu11,alkhalili04, buhr06}
($k_{B}T_{t}=2.8$meV and $k_{B}T_{l}$=45$\mu$eV) have been performed in each case. The corresponding results
are represented in Figure \ref{rate_ani_DR_2} 
for the initial rotational levels $N_{i}^{+}$=0 to 5. 
The differences observed between the various cases enable us to appreciate the sensitivity of the cross sections to the initial rotational level. 
This can be accounted for by the shift in the resonance positions on one hand,  and the rotational coupling between competing molecular states from compatible symmetries on the other hand.

The quantity to needs to computed to compare theory with the
measurements is the average Maxwell anisotropic rate coefficient,
which is obtained by a weighted sum of the Maxwell anisotropic rate
coefficients associated to each of the relevant initial rotational
states - part of them being displayed in Fig.~\ref{rate_ani_DR_2} -
their weights corresponding to the Boltzmann distribution associated
to the assumed rotational temperature.

\begin{figure}
\begin{center}
\includegraphics[width=0.95\columnwidth]{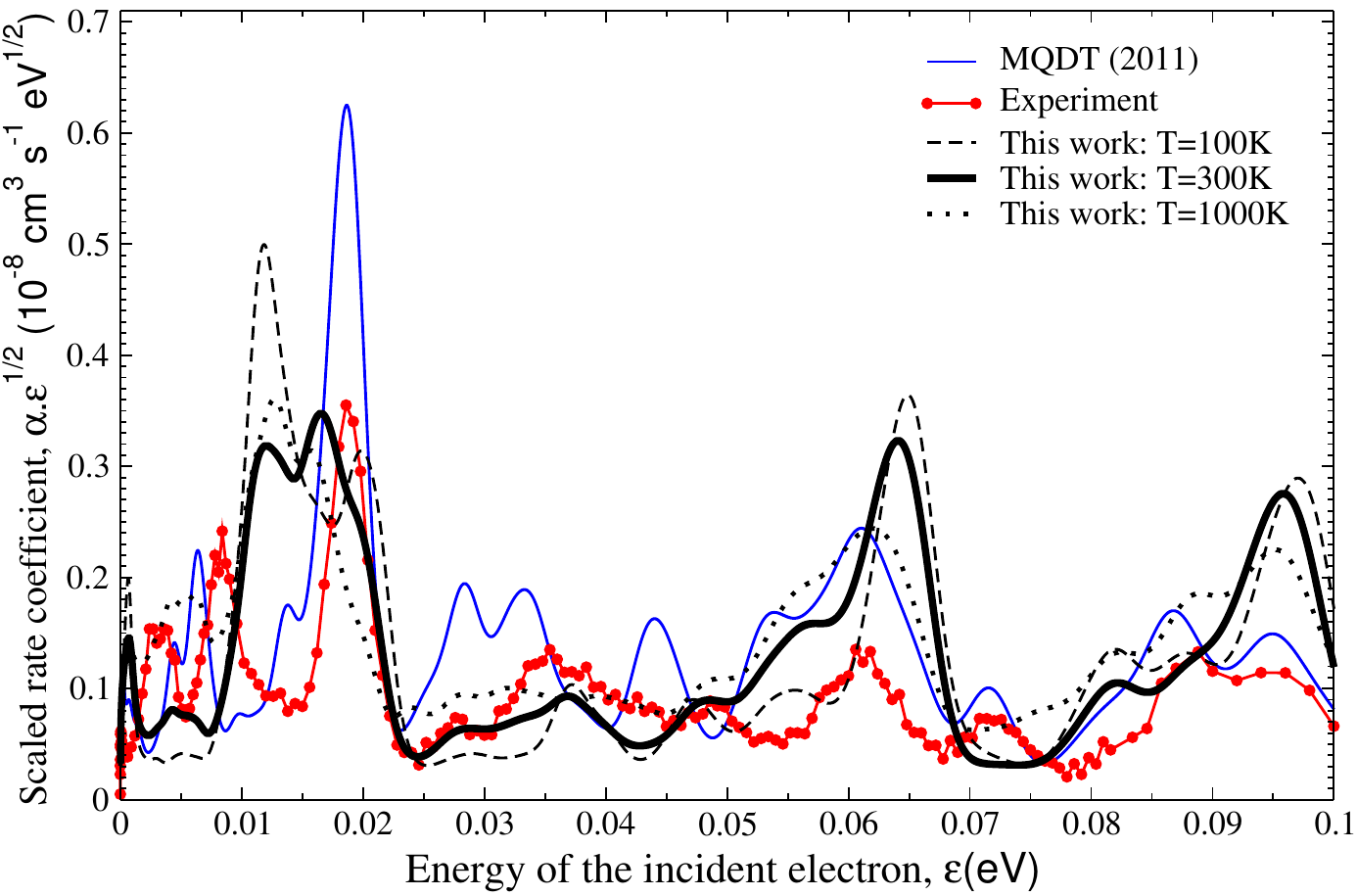}
\end{center}
\caption{\label{cap:rvb3}(Color online) {\textit{Scaled
rate coefficients for the Dissociative Recombination of HD$^{+}(X^{2}\Sigma_{g}^{+})$
at different
experimental
temperatures
assumed for the rotational distribution of the ions in the storage ring, compared to former computations and experiment \cite{waffeu11}.}} }
\end{figure}

The results obtained have been
displayed
in
Fig.~\ref{cap:rvb3} where they are compared with experiment
\cite{alkhalili04, buhr06}
and our former computations
\cite{waffeu11} in which
the only
Rydberg-valence interaction accounted was that characterizing
the $^{1}\Sigma_{g}^{+}$ symmetry.

In Fig.~\ref{cap:rvb3}, differences between our previous theoretical
result - blue curve, \cite{waffeu11} - and the present ones - black
continuous curve - at T=300 K can be noted in terms of positions and
heights of some peaks. This is a consequence of the inclusion of
additional symmetries, as described above, which may induce resonance
shifts through their specific quantum defects, and of the mutual
assistance of compatible symmetries inherent to the rotational
couplings properly accounted for and which causes an attenuation in
the resonant dips.  The prominent peak found by the previous
theoretical approach at 20 meV in agreement with the one detected by
the TSR measurements shifts down to 16 meV, but its amplitude now agrees
 much better with the measured one.  Moreover, in spite of the
appearence of a new prominent peak at almost zero energy and the
increase in the amplitude of that already present close to 100 meV,
the new computed rate coefficient is
closer to the measured one
between 0 and 50 meV.

Since there is some uncertainty in the rotational temperature of the cation in
the storage ring still, we  compute and display in 
Fig.~\ref{cap:rvb3}  the rate coefficient for 100 K and 1000 K,
besides that for 300 K. Indeed, one may notice that for T=1000 K, 
the agreement with experiment is better in terms of cross section amplitudes than that for T=300 K below 10 meV and between 50 and 70 meV.
For T=100 K, 
the agreement is better in the energy range 25-60 meV.
This may mean
that the occupation probability of the rotational levels is not a simple
Maxwell-Boltmann distribution, i.e. temperature
dependent, but includes some dependence on the energy of the incident
electron, perhaps via inelastic and mainly superelastic collisions.

\begin{figure}
\begin{center}
\includegraphics[width=0.95\columnwidth]{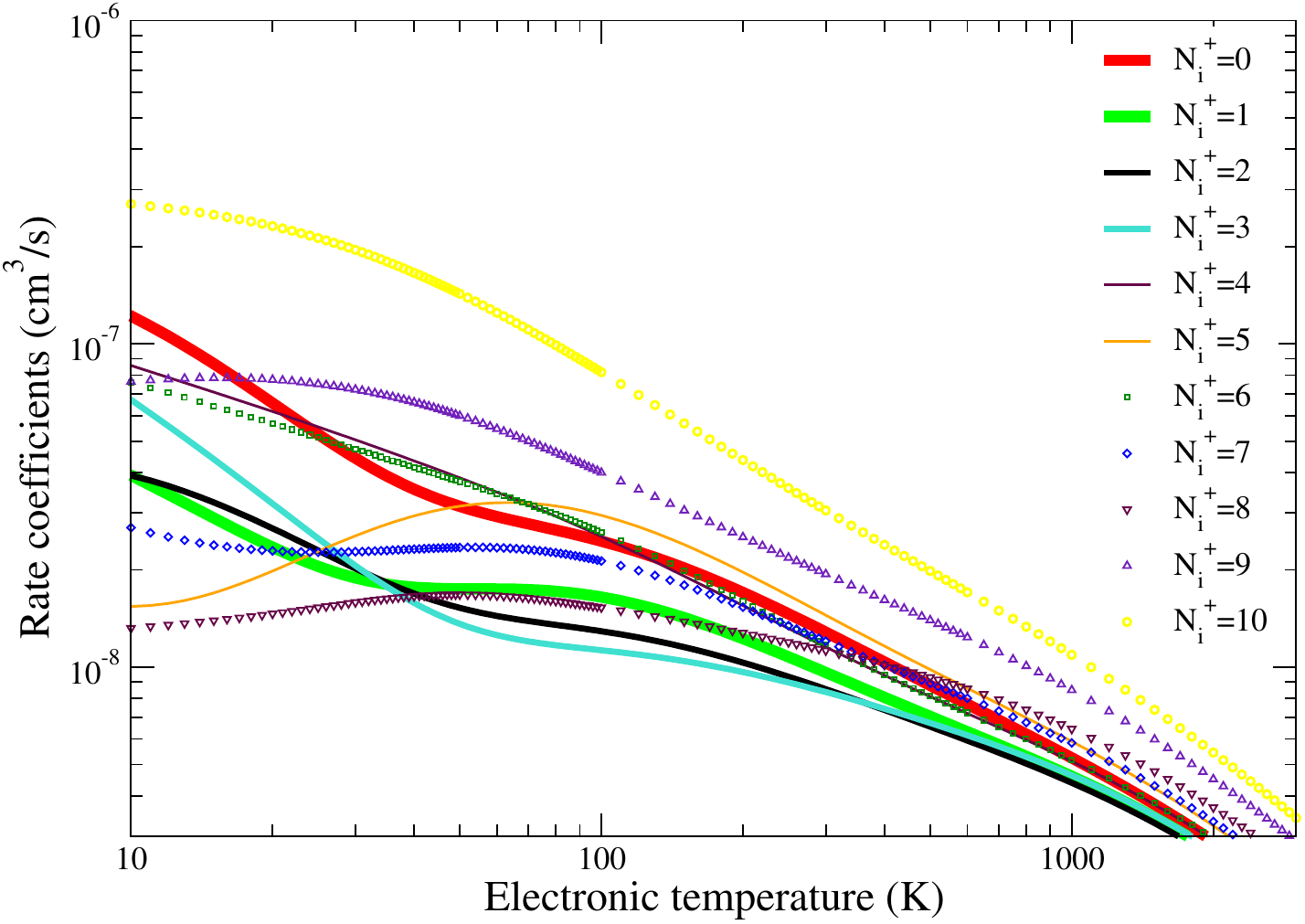}
\end{center}
\caption{\label{rate_iso} (Color online)
{\textit{Maxwell isotropic rate coefficients for the
Dissociative Recombination
HD$^{+}(X^{2}\Sigma_{g}^{+})$ with $v_{i}^{+}=0$ as a function of
initial rotational level, $N_{i}^{+}$=0 to 10
 }.}}
\end{figure}

For further use in
astrochemistry and
cold plasmas modeling, we have performed
{\it isotropic}
Maxwell convolution from all the cross sections
computed. The corresponding isotropic rate coefficients are plotted in Figure \ref{rate_iso} for all the initial
rotational levels. These results show that above the initial rotational state
$N_{i}^{+}$=6, rotational excitation
of the target
favours
dissociative recombination in
the sense that the DR rate coefficients increase with increasing $N_{i}^{+}$.
In contrast, below $N_{i}^{+}$=6 rotational excitation
slows down DR.

\section{\textbf{Conclusion}}

 A new series of computations have been carried out for the collisions between HD$^{+}$ molecular ions and
 electrons with kinetic energy below 300 meV.
 Whereas in our previous calculations on this species \cite{waffeu11} the only Rydberg-valence
 interactions considered were those within the dominant symmetry $^{1}\Sigma_{g}^{+}$,
in the present ones, this symmetry contributes simultaneously with the '+' components
of the 
  $^{1}\Pi_{g}$ and $^{1}\Delta_{g}$ symmetries, the corresponding
 states being rotationally coupled. Moreover, we have involved in the
 same way in the present approach the triplet gerade, and the singlet
 and triplet ungerade symmetries. The account of these numerous states
 and interactions makes the current treatment more accurate than the
 previous ones.  A further step in the same direction will be the use
 of more accurate molecular data. This will be achieved by using the
 latest and ongoing progress in the R-matrix and MQDT techniques
 ~\cite{little_2014, argoubi_2011, oueslati_2014} in producing
 potential energy curves and electronic couplings, which can be used
tp produce realistic models (positions, amplitudes and widths)
 for the Rydberg resonances dominating the shape of the cross sections.
 An alternative route to improved accuracy is the use of the global
 version of the MQDT~\cite{jungen_2011}, which has proved its
 efficiency in the spectroscopy of the neutral  H$_2$ and
 its isotopes~\cite{mezei_2012}.  

Cross sections, Maxwell anisotropic and isotropic rate coefficients have been obtained for state-to-state
 rotational transitions induced in HD$^{+}$ molecular ions by electronic collisions.  To the best of our
 knowledge, this is the first ever reported result on this process, using the MQDT-based method. At very low
 collision energy, the
Maxwell anisotropic rate coefficients
computed with both MQDT and ANR/R-Matrix methods
are in good agreement. They are comparable to within 30 percent with
those resulting from a fit of the experimental cooling curves, derived from the
 the measurements performed at the Heidelberg Test Storage Ring~\cite{schwalm}.
These latter results account for the strong contribution of the rotational
de-excitation to the rotational cooling of HD$^{+}$ molecular ions, as reported
by Shafir {\it et al.}~\cite{shafir} and Schwalm {\it et al.}~\cite{schwalm}.

Furthermore, new cross sections, Maxwell anisotropic and isotropic rate coefficients
have been derived for the Dissociative Recombination of HD$^{+}$ molecular ions.
In spite of the displacement of few prominent resonances with respect to the previous predicted positions and$/$or with respect to the positions found by the measurements, the agreement
between theory and experiment has been improved at low energy.
The numerical data
for all the initial $N_{i}^{+}$=0 to 10, ready to be used in the kinetics modeling in astrochemistry and cold plasma physics,
are available upon request.  

Similar computations are
ongoing
for the H$_{2}^{+}$ and D$_{2}^{+}$ molecular ions, and the corresponding results will be reported in a future article.

\begin{acknowledgments}
The authors thank A. Wolf, D. Schwalm, D. Dowek, Ch. Jungen, H. Takagi and S. L. Guberman for numerous discussions,
and S. Ilie for technical assistance.

O.M. is greatful to the International Atomic Energy
Agency (IAEA, Vienna) for financial support through the Contract No 16712 with the University of Douala
(CAMEROON), and to the Laboratoire Ondes et Milieux Complexes of
University of Le Havre for hospitality.
  
The authors acknowledge scientific and financial support from the IAEA,  via the Coordinated Research Projects
"Atomic and Molecular Data for State-Resolved Modelling of Hydrogen and Helium and their Isotopes in   Fusion Plasmas", 
and "Light Element Atom, Molecule and Radical Behaviour
in the Divertor and Edge Plasma Regions"), 
as well as from the European Spatial Agency (ESA) via the contract ESTEC 21790/08/NL/HE.

We are grateful for the financial support from the 
French Research Federation for Fusion Studies (CEA, EFDA, EURATOM), 
from Agence Nationale de la Recherche via the projects 
'SUMOSTAI' (ANR-09-BLAN-020901)
and 
'HYDRIDES' (ANR-12-BS05-0011-01), 
from the 
IFRAF-Triangle de la Physique via the project 'SpecoRyd' , 
and from the 
CNRS via the programme `Physique et Chimie du Milieu Interstellaire', 
and
the CNRS PEPS projects `Physique th\'{e}orique et ses interfaces'
TheMS and TPCECAM. 

We thank for generous financial support from La R\'egion Haute-Normandie via the CPER 'THETE' project,
and the GRR Electronique, Energie et Matériaux.

Part of this work has been performed in the frame of the "F\'ed\'eration de Recherche Energie, Propulsion, Environnement",
and of the LabEx EMC$^3$, via the project PicoLIBS (ANR-10-LABX-09-01).

I.F.S. thanks the Laboratoire   Aim\'e Cotton for hospitality.
\end{acknowledgments}
%

\section*{Data availability}
Upon a reasonable request, the data supporting this article will be provided by the corresponding author.


\begin{thebibliography}{99}
\bibitem{coppola} C. M. Coppola, S. Longo, M. Capitelli, F. Palla, and D. Galli, Astrophys. J. Suppl. Ser. {\bf{193}}, 7 (2011).

\bibitem{roueff} F. Le Petit, E. Roueff, and J. Le Bourlot,
Astronomy and Astrophysics {\bf{390}}, 369 (2002);
E. Roueff and F. Le Petit, in {\em Astrochemistry: Recent
Successes and Current Challenges, Proceedings of the IAU Symposium No 231}, edited by  D. C. Lis, G. A. Blake, and E. Herbst (Cambridge University Press, Cambridge, 2006), p. 197;
 M. Ag\'undez, J.R.
Goicochea, J. Cernicharo, A. Faure, and E. Roueff,
Astrophys. J. {\bf{713}}, 662 (2010).

\bibitem{black} J. H. Black and A. Dalgarno, Astrophys. J. {\bf{203}}, 132 (1976); 
E. F. Van Dishoeck and J. H. Black, Astrophys. J. Suppl. Ser. {\bf{62}}, 109 (1986); 
J. H. Black and E. F. Van Dishoeck, Astrophys. J. {\bf{322}}, 412 (1987).

\bibitem{gay} C. D. Gay, N. P. Abel, R. L. Porter, P. C. Stancil, G. J. Ferland, G. Shaw, P. A. M. Van Hoof, and R. J. R. Williams,
Astrophys. J. {\bf{746}}, 78 (2012).

\bibitem{molscat} J. M. Hutson and S. Green, MOLSCAT computer code, version 14 (UK: Collaborative
 Computation Project No 6 of the Science and Engineering Council), 1994.

\bibitem{rabadan98b} I. Rabad\'an, B. K. Sarpal, and J. Tennyson,  Mon. Not. R. astr. Soc., {\bf{299}}, 171 (1998).

\bibitem{lim99} A.J. Lim, I. Rab\'adan and J. Tennyson,
Mon. Not. R. astr. Soc., {\bf 306}, 473 (1999).

\bibitem{faure01} A. Faure and J. Tennyson, Mon. Not. R. astr. Soc.,   {\bf{325}}, 443 (2001).

\bibitem{JS06} I. Jimenez-Serra, J. Martin-Pintado, S. Viti, S. Martin, A. Rodriguez-Franco,
A. Faure and J. Tennyson, Astrophys. J., {\bf 650}, L135 (2006).

\bibitem{faure02} A. Faure and J. Tennyson, J. Phys. B {\bf{35}}, 3945 (2002).

\bibitem{faure03} A. Faure and J. Tennyson, Mon. Not. Roy. astr. Soc., {\bf 340}, 468 (2003).

\bibitem{faure06} A. Faure, V. Kokoouline, C. H. Greene, and J. Tennyson, J. Phys. B {\bf{39}}, 4261 (2006).

\bibitem{kokoo10} V. Kokoouline, A. Faure, J. Tennyson, and C. H. Greene,  Mon. Not. R. astr. Soc., {\bf{405}}, 1195 (2010).

\bibitem{jt474} J. Tennyson, Phys. Rep., {\bf 491}, 29 (2010).

\bibitem{rabadan98a} I. Rabad\'an and J. Tennyson, Comput. Phys. Commun. {\bf{114}}, 129 (1998).

\bibitem{rawlings} J. M. C. Rawlings, J. E. Drew, and M. J. Barlow,  Mon. Not. R. astr. Soc., {\bf{265}}, 968 (1993).

\bibitem{janev} R. T. Janev, {\em Atomic and Molecular Processes in Fusion Edge
Plasmas} (Plenum, New York, 1995).

\bibitem{reiter} D. Reiter, in {\em Nuclear
Fusion Research: Understanding Plasma Surface Interaction} edited by R. E. H. Clarke and D. Reiter (Springer, Heidelberg,
2005) p. 29.

\bibitem{fantz} U. Fantz, D. Reiter, B. Heger, and D. Coster, J. Nucl. Mat. {\bf{290-293}}, 367 (2001); 
U. Fantz and P. T. Greenland, Contrib. Plasma Phys. {\bf{42}}, 694 (2002); 
U. Fantz and D. Wunderlich, in {\it Atomic and Plasma-Material Interaction Data for Fusion}, Vol. 14 (International Atomic Energy Agency, Vienna, 2008), p. 56; 
U. Fantz and D. Wunderlich, in {\it Atomic and Plasma-Material Interaction Data for Fusion}, Vol. 16 (International Atomic Energy Agency, Vienna, 2011), p. 76.

\bibitem{zhaunerchyk07} V. Zhaunerchyk, A. Al-Khalili, R. D. Thomas, W. D. Geppert, V.
Bednarska, A. Petrignani, A. Ehlerding, M. Hamberg, M. Larsson, S.
Rosen, and W. J. van der Zande, Phys. Rev. Lett. \textbf{99}, 013201 (2007).


\bibitem{deruette07} N. de Ruette, Ph.D. thesis, Universit\'{e} Catholique de Louvain la Neuve, 2007.

\bibitem{mot08} O. Motapon, F. O. Waffeu Tamo, X. Urbain, and I. F. Schneider, Phys. Rev. A \textbf{77}, 052711 (2008).

\bibitem{zajfman93} P. Forck, M. Grieser, D. Habs, A. Lampert, R. Repnow, D. Schwalm, A. Wolf, and D. Zajfman, Phys. Rev. Lett {\bf 70}, 426 (1993).

\bibitem{takagi95} T. Tanabe, I. Katayama, H. Kamegaya, K. Chida, Y. Arakaki, T. Watanabe, M. Yoshizawa, M. Saito, Y. Haruyama, K. Hosono, K. Hatanaka, T. Honma, K. Noda, S. Ohtani, and H. Takagi, Phys. Rev. Lett. {\bf 75}, 1066 (1995).

\bibitem{stromholm95} C. Str{\"o}mholm, I.~F. Schneider, G. Sundstr{\"o}m, L. Carata, H. Danared, S. Datz,
O. Dulieu, A. K{\"a}llberg, M. af Ugglas, X. Urbain, V. Zengin, A. Suzor-Weiner, and M. Larsson, Phys. Rev. A {\bf 52}, R4320 (1995).

\bibitem{andersen95} J. R. Mowat, H. Danared, G. Sundstr{\"o}m, M. Carlson, L. H. Andersen, L.
Vejby-Christensen, M. af Ugglas, and M. Larsson, Phys. Rev. Lett. {\bf 74}, 50 (1995).

\bibitem{ifs-a18} I.~F. Schneider, C. Str{\"o}mholm, L. Carata, X. Urbain, M. Larsson,  and A. Suzor-Weiner, J. Phys. B {\bf 30}, 2687 (1997).

\bibitem{amitay98} Z. Amitay, A. Baer, M. Dahan, L. Knoll, M. Lange, J. Levin, I. F. Schneider, D. Schwalm, A. Suzor-Weiner, Z. Vager, R.Wester, A. Wolf, and D. Zajfman, Science \textbf{281}, 75 (1998).

\bibitem{amitay99} Z. Amitay, A. Baer, M. Dahan, J. Levin, Z. Vager, D. Zajfman, L. Knoll, M. Lange, D. Schwalm, R. Wester, A. Wolf , I.~F. Schneider, and A. Suzor-Weiner, Phys. Rev. A {\bf{60}}, 3769 (1999).

\bibitem{takagi03} H. Takagi, Physica Scripta \textbf{T96}, 52 (2002).

\bibitem{alkhalili04} A. Al-Kalili \emph{et al}., Phys. Rev. A {\bf 68}, 042702 (2003).

\bibitem{buhr06} H. Buhr, Ph.D. thesis, University of Heidelberg, 2006.

\bibitem{waffeu11} F. O. Waffeu Tamo, H. Buhr, O. Motapon, S. Altevogt, V. M. Andrianarijaona, M. Grieser, L. Lammich,
M. Lestinsky, M. Motsch, I. Nevo, S. Novotny, D. A. Orlov, H. B. Pedersen, D. Schwalm, F. Sprenger,
X. Urbain, U. Weigel, A. Wolf, and I. F. Schneider, Phys. Rev. A \textbf{84}, 022710 (2011).

\bibitem{shafir} D. Shafir et {\em al}., Phys. Rev. Lett. {\bf{102}}, 223202 (2009).

\bibitem{schwalm} D. Schwalm et {\em al}., J. Phys.: Conf. Ser. {\bf{300}}, 012006 (2011).

\bibitem{annick80} A. Giusti-Suzor,  J. Phys. B {\bf 13}, 3867 (1980).

\bibitem{ja77} Ch. Jungen  and O. Atabek, J. Chem. Phys. {\bf 66}, 5584 (1977).

\bibitem{bardsley68} N. Bardsley, J. Phys. B {\bf 1}, 349 (1968); {\bf 1}, 365 (1968).

\bibitem{sidis72} V. Sidis  and H. Lefebvre-Brion,  J. Phys. B {\bf 4}, 1040 (1971).

\bibitem{ngassam03b} V. Ngassam, A. Florescu, L. Pichl, I. F. Schneider, O. Motapon, and A. Suzor-Weiner, Eur. Phys. J. D \textbf{26}, 165 (2003).

\bibitem{valcu} B. V\^alcu, I. F. Schneider, M. Raoult, C. Str{\"o}mholm, M. Larsson, and A. Suzor-Weiner, Eur. Phys. J. D {\bf 1}, 71 (1998).

\bibitem{fano70} U. Fano, Phys. Rev. A {\bf{2}}, 353 (1970).

\bibitem{chfan} E.~S. Chang and U.~Fano, Phys. Rev. A {\bf 6}, 173 (1972).

\bibitem{seaton83} M.~J. Seaton,  Rep. Prog. Phys. {\bf 46}, 167 (1983).

\bibitem{tennyson96} J. Tennyson, At. Data. Nucl. Data Tables {\bf 64}, 253 (1996).

\bibitem{TJ} M. Telmini and Ch. Jungen, Phys. Rev. A \textbf{68}, 062704 (2003).

\bibitem{S-ABOTJ} F. Argoubi, S. Bezzaouia, H. Oueslati, M. Telmini, and Ch. Jungen, Phys. Rev. A \textbf{83}, 052504 (2011).

\bibitem{BATJ} S. Bezzaouia, F. Argoubi, M. Telmini, and Ch. Jungen, J. Phys.: Conf. Ser. \textbf{300}, 012013 (2011).

\bibitem{BT} S. Bezzaouia and M. Telmini, AIP Conf. Proc. \textbf{935}, 183 (2007).

\bibitem{tbj} M. Telmini, S. Bezzaouia, and Ch. Jungen, Int. Journal. Quant. Chem. \textbf{104}, 530 (2005).

\bibitem{BTJ} S. Bezzaouia, M. Telmini, and Ch. Jungen, Phys. Rev. A \textbf{70}, 012713 (2004).

\bibitem{wol3} W. Kolos and L. Wolniewicz, J. Chem. Phys. \textbf{50}, 3228 (1969); 
L. Wolniewicz and K. Dressler, J. Chem. Phys. \textbf{100}, 444 (1994); 
T. Orlikowski, G. Staszewska, and L. Wolniewicz, Mol. Phys. \textbf{96}, 1445 (1999).

\bibitem{lim diss} T. E. Sharp, At. Data. Nucl. Data Tables \textbf{2}, 119 (1970); 
I. Sanchez and F. Mart\'in, J. Chem. Phys. \textbf{106}, 7720 (1997).

\bibitem{little_2014} D. A. Little, and J. Tennyson, J. Phys. B \textbf{47}, 105204 (2014).
\bibitem{argoubi_2011} F. Argoubi, S. Bezzaouia, H. Oueslati, M. Telmini, and Ch. Jungen,
Phys. Rev. A \textbf{83}, 052504 (2011).
\bibitem{oueslati_2014} H. Oueslati, F. Argoubi, S. Bezzaouia, M. Telmini, and Ch. Jungen,
Phys. Rev. A \textbf{89}, 032501 (2014).
\bibitem{jungen_2011}Ch. Jungen, in {\it Elements of Quantum Defect Theory}, Handbook of 
High-resolution Spectroscopy, edited by M. Quack and F. Merkt (Wiley $\&$ Sons, New York, 2011), p. 471.
\bibitem{mezei_2012} J. Zs. Mezei, I. F. Schneider, E. Roueff, and Ch. Jungen, Phys. Rev. A \textbf{85}, 043411 (2012).
\end{thebibliography}
\end{document}